
  \magnification\magstep2

\baselineskip = 0.65 true cm
\parskip=0.5 true cm
                           
  \def\sa{\vskip 0.30 true cm}
  \def\sb{\vskip 0.60 true cm}



\def\grt{\hbox{\bf R$^3$}}
\def\grq{\hbox{\bf R$^4$}}
\def\grc{\hbox{\bf R$^5$}}
\def\grn{\hbox{\bf N}}

\def\gr{\hbox{\bf R}}

\rightline{\hbox{\bf LYCEN 9109}}
\rightline{April 1991}
\rightline{Revised: June 1991}

\sa
\sb
\sa

\noindent {\bf On the $q$-analogue of the hydrogen atom}

\sa
\sb

\noindent M Kibler$^1$ and T N\'egadi$^2$ 

\sa

\noindent $^1$Institut de Physique Nucl\'eaire de Lyon, 
IN2P3-CNRS et Universit\'e Claude Bernard, 
F-69622 Villeurbanne Cedex, France 

\sa

\noindent $^2$Laboratoire de Physique Th\'eorique, 
Institut des Sciences Exactes, 
Universit\'e d'Oran, 
Es-S\'enia, Oran, Alg\'erie 

\sa
\sb
\sb


\sa

\noindent {\bf Abstract}. The discrete spectrum of a 
$q$-analogue of the hydrogen atom is obtained from a
deformation of the Pauli equations. As an alternative, the 
spectrum is derived from a deformation of the four-dimensional 
oscillator arising in the application of the 
Kustaanheimo-Stiefel transformation to the hydrogen atom. A 
model of the $2s-2p$ Dirac shift is proposed in the context of 
$q$-deformations. 

\sa
\sb
\sa
\sb
\sb
\sa
\sb
\sb

\noindent (published in Journal of Physics A: 
Mathematical and General 24 (1991) 5283-5289)

\vfill\eject
\baselineskip = 0.84 true cm

\noindent {\bf 1. Introduction} 

\noindent In recent years, 
a great deal of work has been devoted to 
an apparently new mathematical structure, the structure of 
quantum group, and its application to various fields of physics 
(statistical mechanics, conformal quantum field theory, etc.). Indeed, 
the structure of quantum group (or algebra) is connected to 
those of Hecke algebra and quasi-triangular Hopf algebra and 
certainly plays an important r\^ole in non-commutative geometry 
(Drinfel'd 1985, Jimbo 1985, Woronowicz 1987). 
Loosely speaking, the structure of quantum algebra corresponds 
to a deformation which can be characterised by a deformation 
parameter $q$, an arbitrary complex number in the most general 
case. 

\smallskip

There is no universal significance of the 
parameter $q$ and the physicist is tempted to consider it as a 
phenomenological parameter, something like a curvature constant, 
to be adjusted to experimental data. The limiting situation $q = 1$ 
corresponds to the flat case and gives back the results 
afforded by (ordinary) quantum mechanics. For example, when $q$ 
goes to 1, the quantum algebra $su_q(2)$, which defines a 
$q$-analogue of angular momentum, reduces to the Lie algebra 
$su(2)$ of (ordinary) angular momentum. The algebra $su(2)$ is 
used from atoms to quarks with some reasonable success and, 
therefore, the replacement of $su(2)$ by $su_q(2)$ with $q$ 
near to 1 should be appropriate for the description of fine 
structure effects. In the framework of this philosophy, we may 
expect to obtain a fine structure of the hydrogen atom from a 
deformation of its non-relativistic spectrum.

\smallskip

It seems interesting to investigate $q$-analogues of 
dynamical systems in view of their potential use as refined 
units for modeling physical systems, with the case $q-1 = 0$ 
corresponding to already known effects and the case 
$q-1 \approx 0$ to new effects (as for instance 
spectrum shift and spectrum splitting). In this vein, the 
$q$-analogue of the harmonic oscillator, as derived among 
others by Macfarlane (1989), Biedenharn (1989), Sun and Fu 
(1989), and Kulish and Reshetikhin (1989), 
represents a first important step ; further, the $q$-analogues 
of various coherent states introduced recently (Quesne 1991, 
Katriel and Solomon 1991) might be interesting in physical 
applications.

\smallskip

It is the purpose of the present work to introduce a 
$q$-analogue of another very simple dynamical system, viz, the 
three-dimensional hydrogen atom. We shall deal here with the (quantum 
mechanical) discrete spectrum of the hydrogen atom in 
$\gr^3$ and shall obtain in \S 2 its $q$-analogue by passing from 
the dynamical invariance algebra $so(4) = su(2) \oplus su(2)$ 
of the ordinary hydrogen atom system to its $q$-analogue 
$su_q(2) \oplus su_q(2)$. In \S 3, an alternative spectrum will
be derived from the Kustaanheimo-Stiefel (KS) transformation 
(Kustaanheimo and Stiefel 1965) and applied to a 
phenomenological derivation of the $2s-2p$ fine structure splitting. 
We shall close this paper with some conclusions in \S 4 about 
the non-unicity of $q$-analogues and the relevance of the 
quantum algebra $so_q(3,2)$ for the Wigner-Racah algebra of 
$SU_q(2)$. 

\smallskip

\noindent {\bf 2. $q$-analogue of the Pauli equations}

\noindent Since we want to examine $su_q(2) \oplus su_q(2)$, we 
review, in a non-standard presentation, some basic facts 
about the quantum algebra $su_q(2)$. Let 
$ {\cal E} \; = \; \left\{ |jm> \ : 2j \in \grn, \ m = -j(1)j \right\} $ 
be the Hilbert space of the representation theory of 
the (ordinary) Lie group $SU(2)$. We define the 
operators $a_+$,    $a_+^+$, 
          $a_-$ and $a_-^+$ acting on ${\cal E}$ by the relations
$$
a_{\pm}|jm> \; = \; 
{\sqrt{[j \pm m]}} \; |j - {1\over 2}, m \mp {1\over 2}>
$$
$$
\eqno (1)
$$
$$
a_{\pm}^+|jm> \; = \; 
{\sqrt{[j \pm m +1]}} \; |j + {1\over 2}, m \pm {1\over 2}>.
$$
In equation (1), we use the notation $\left[ x \right] \equiv 
                                 \left[ x \right]_q$ where 
$$
[x]_q \; = \; { {q^x - q^{-x}} \over {q - q^{-1}} }
\eqno (2)
$$
is the $q$-analogue of the (real) number $x$ ; in this paper, $q$ 
is taken as a complex number which is not a root of 
unity. In the limiting situation $q = 1$, equation (1) gives back the 
defining relations introduced by Schwinger (1952) in his theory 
of angular momentum (see also Kibler and Grenet (1980) and 
Nomura (1990)). Some trivial properties (matrix elements, 
time-reversal behavior, adjointness relations, etc.) for the 
operators $a_+$, $a_+^+$, $a_-$ and $a_-^+$ can be derived from 
the starting relations (1). 

\smallskip

By introducing
$$
n_1 \; = \; j + m \qquad \quad 
n_2 \; = \; j - m \qquad \quad 
|jm> \; \equiv \; |j + m, j - m> \; = \; |n_1,n_2>  \eqno (3)
$$
we immediately obtain
$$
  a_+|n_1n_2>   \; = \; {\sqrt {[n_1]}} \; |n_1 - 1, n_2> 
\qquad 
  a_+^+|n_1n_2> \; = \; {\sqrt {[n_1 + 1]}} \; |n_1 + 1, n_2> 
$$
$$
\eqno (4)
$$
$$
  a_-|n_1n_2>    = {\sqrt {[n_2]}} \; |n_1, n_2 - 1> 
\quad 
  a_-^+|n_1,n_2> = {\sqrt {[n_2 + 1]}} \; |n_1, n_2 + 1>. 
$$
Consequently, the sets $\left\{ a_+, a_+^+ \right\}$ 
                   and $\left\{ a_-, a_-^+ \right\}$ are two commuting sets of 
$q$-bosons. Indeed, these $q$-bosons satisfy the relations
$$
a_+a^+_+ \; - \; q^{-1} a^+_+ a_+ \; = \; q^{N_1} \qquad \quad 
a_-a^+_- \; - \; q^{-1} a^+_- a_- \; = \; q^{N_2}
$$
$$
\eqno (5)
$$
$$
[a_+, a_-]_- \; = \; [a^+_+, a^+_-]_- \; = \; 
[a_+, a^+_-]_- \; = \; [a^+_+, a_-]_- \; = \; 0
$$
where $N_1$ and $N_2$ are the (usual) number operators defined by
$$
N_1 |n_1n_2 > \; = \; n_1 |n_1n_2> \qquad \quad 
N_2 |n_1n_2 > \; = \; n_2 |n_1n_2>. 
\eqno (6)
$$
Equations (4-6) show that each of the sets 
$\left\{ a_+, a_+^+ \right\}$ and 
$\left\{ a_-, a_-^+ \right\}$ is a set of 
$q$-bosons as introduced independently by many authors including 
Macfarlane (1989), Biedenharn (1989), and Sun and Fu (1989). 

\smallskip

We now introduce the three following operators
$$
J_- \; = \; a^+_- a_+ \qquad \quad 
J_3 \; = \; {1\over 2} \left( N_1 - N_2 \right) \qquad \quad
J_+ \; = \; a^+_+ a_-. 
\eqno (7)
$$
The action of $J_-$, $J_3$ and $J_+$ on the space ${\cal E}$ is given by
$$
    J_3 |jm> \; = \; m \; |jm> \qquad \quad 
J_{\pm} |jm> \; = \; 
{\sqrt {[j \mp m] [j \pm m + 1]}} \; |j, m \pm 1>.
\eqno (8)
$$
We thus recover the Jimbo representation of the quantum group 
$SU_q(2)$. The commutation relations of the operators $J_-$, 
$J_3$ and $J_+$ easily follow from (8). As a matter of fact, we 
have
$$
[J_3, J_{\pm}]_- \; = \; \pm \; J_{\pm} \qquad \quad
[J_+, J_-]_- \; = \; [2J_3]. 
\eqno (9)
$$
In equation (9), the abbreviation $[A] \equiv [A]_q$ stands for 
the $q$-analogue 
$$
[A]_q \; = \; { {q^A - q^{-A}} \over {q - q^{-1}} }
\eqno (10)
$$
of the operator $A$ acting on ${\cal E}$. The relations (9) 
characterise the quantum algebra $su_q(2)$. 

\smallskip

A basic ingredient 
of $su_q(2)$, of central importance for what follows, is its Casimir 
operator $J^2$. We shall take it in the form
$$
J^2 \; = \; {1\over 2} (J_+J_- \; + \; J_-J_+) \; + \; 
{ {[2]} \over {2} } \; [J_3]^2
\eqno (11)
$$
so that the eigenvalues of $J^2$ on ${\cal E}$ are 
$[j] \left[ j+1 \right]$ with $2j$ in $\grn$. The form taken in 
the present paper for $J^2$ differs from the one usually 
encountered in the literature ; indeed, equation (11) ensures 
that the eigenvalues of $J^2$ are merely $j(j+1)$ in the 
limiting case $q=1$, a result which turns out to be essential 
in the framework of the philosophy sketched in the 
introduction.

\smallskip

We are in a position to $q$-deform the Pauli equations for 
the three-dimensional hydrogen atom in the case of the discrete 
spectrum. From the orbital angular momentum {\bf L} and the 
Laplace-Runge-Lenz-Pauli vector {\bf M} for the ordinary hydrogen atom 
system, we define the operators {\bf A} and {\bf B} through
$$
{\bf A} \; = \; {1\over 2} ({\bf L} + {\bf N}) \qquad \quad
{\bf B} \; = \; {1\over 2} ({\bf L} - {\bf N}) \qquad \quad
{\bf N} \; = \; \sqrt{ {-\mu} \over {2E} } \; {\bf M}
\eqno (12)
$$
where $E$ is the energy ($E < 0$) and $\mu$ the reduced mass of 
the atom. We know that 
$\left\{ A_i \ : \ i = 1,2,3 \right\}$ and 
$\left\{ B_i \ : \ i = 1,2,3 \right\}$ are the generators of two groups, say 
$ASU(2)$ and $BSU(2)$, both of them being of the $SU(2)$ type.
In addition, the Casimir operators ${A}^2$ and 
                                   ${B}^2$ of the Lie algebras 
$asu(2)$ and $bsu(2)$ satisfy the equations (see Pauli 1926)
$$
A^2 - B^2 \; = \; 0 \qquad \quad 
E \, \left( 2A^2 + 2B^2 + \hbar^2 \right) \; 
= \; - {1\over 2} \; \mu \; Z^2 \; e^4
\eqno (13)
$$
valid for a hydrogenlike atom with a nuclear charge $Ze$. 
A possible way to define a $q$-deformed hydrogen atom is to 
deform the Lie groups $ASU(2)$ and $BSU(2)$. Then, the Pauli 
equations (13) have to be extended to the quantum algebra 
$asu_q(2) \oplus bsu_q(2)$. This deformation produces a 
$q$-analogue of the hydrogen atom, the energy $E$ of which is given by
$$
E \; \equiv \; E_j \; = \; { {1} \over {4[j] [j+1] + 1} } \; E_{0} 
\qquad \quad 2j \in \grn 
\eqno (14)
$$
with
$$
E_{0} \; = \; - \; {1\over 2} \; { {\mu \; Z^2 \; e^4} \over {\hbar^2} }. 
\eqno (15)
$$
Equations (14) and (15) follow from (13) where we have 
introduced the eigenvalues of $A^2$ and $B^2$, in correct 
units, of the Casimir operators of $asu_q(2)$ and $bsu_q(2)$.

\smallskip

The quantity $E_0$ is the energy of the ground state of the 
$q$-deformed hydrogen atom. The $q$-deformed hydrogen atom 
defined by (14) thus has 
the same ground energy level as the ordinary hydrogen 
atom which corresponds to the limiting situation $q=1$. 
Furthermore, its (discrete) spectrum exhibits the same degeneracy 
as that of the ordinary hydrogen atom. The only difference 
between the cases $q = 1$ and $q \ne 1$ arises in the position 
of the excited levels. Of course, the $q$-deformed spectrum 
reduces to that of the ordinary hydrogen atom when $q$ goes 
to 1 ; the principal quantum number $n$ is then given by 
$n = 2j + 1$. 

\smallskip

\noindent {\bf 3. $q$-analogue and KS transformation}

\noindent In the case $q=1$, we know how to pass from the 
hydrogen atom in $\grt$ to an isotropic harmonic oscillator in 
$\grq$ or $\gr^2 \otimes \gr^2$ by applying the 
KS transformation (Boiteux 1972, Kibler 
and N\'egadi 1983a,~b, 1984, Kibler, Ronveaux and N\'egadi 
1986). The reader may consult Kustaanheimo and Stiefel (1965), 
Cornish (1984), Lambert and Kibler (1988), and Hage Hassan and 
Kibler (1991) for a description of this transformation, which 
is indeed a particular Hurwitz transformation. The KS 
transformation can be used to define a $q$-deformed hydrogen 
atom. This may be achieved along the following lines : (i) 
apply the KS transformation to the ordinary hydrogen 
atom in $\gr^3$ in order to obtain an isotropic harmonic oscillator 
     in $\gr^4$, (ii) transpose the latter oscillator 
into its $q$-analogue and (iii) invoke the ``inverse'' 
KS transformation to get a $q$-analogue of the hydrogen atom. 

\smallskip

As a result, we obtain a $q$-deformed hydrogen atom 
characterised by the discrete spectrum
$$
E \equiv E_{n_1n_2n_3n_4} \; = 
                          \; { {16} \over {\nu(n_1n_2n_3n_4)^2} } \; E_{0}
\eqno (16a)
$$
$$
\nu(n_1n_2n_3n_4) 
\; = \; \sum_{i = 1}^{4} \; \left( [n_i] + [n_i + 1] \right) 
\qquad n_i \in \grn 
\qquad (i = 1,2,3,4). 
\eqno (16b)
$$
In the limiting case $q=1$, the principal quantum number $n$ is 
connected to the oscillator quantum numbers $n_i$ 
($i = 1, 2, 3, 4$) by $n_1 + n_2 + n_3 + n_4 + 2 = 2n$. Then, 
the $q$-deformed spectrum (16) gives back the discrete spectrum 
of the ordinary  hydrogen atom when $q$ goes to 1. Here again, 
the $q$-deformed hydrogen atom has the same ground energy level 
as the ordinary hydrogen atom. However, we now pass from the 
discrete spectrum of the ordinary hydrogen atom to the one of 
this alternative $q$-deformed hydrogen atom by means of a level 
splitting for the excited levels.

\smallskip 

As an illustration, the level corresponding to the principal quantum 
number $n = 2$ when $q = 1$ is split into two 
levels when $q \ne 1$. In fact, we have the two energy levels
$$
E_{2000} = {16 \over ([2] + [3] + 3)^2 } \; E_0 \quad \qquad
E_{1100} = { 4 \over ([2] + 2)^2       } \; E_0.
\eqno (17)
$$
We thus have a splitting of the $n=2$ non-relativistic level, 
a situation which is reminiscent of the fine structure 
splitting afforded by the Dirac theory of the hydrogen atom. 
By using the oscillator basis $ \left \{ \Phi_{n_1n_2n_3n_4} \right \} $ 
described by Kibler, Ronveaux and 
N\'egadi (1986), it can be proved that the wave functions 
$\Psi_{n \ell m}$ corresponding to the limiting case $q=1$ are 
$$
\eqalign{
\Psi_{200} & = N \left( \Phi_{2000} + \Phi_{0200}+\Phi_{0020}+\Phi_{0002} \right) 
\cr 
\Psi_{210} & = N \left( \Phi_{2000} + \Phi_{0200}-\Phi_{0020}-\Phi_{0002} \right) 
} \eqno (18)
$$
for (the doublet) $E_{2000}$, and 
$$
\eqalign{
\Psi_{211}  & = N \left[ \Phi_{1010} - \Phi_{0101} +
                i \left( \Phi_{1001} - \Phi_{0110} \right) \right]
\cr 
\Psi_{21-1} & = N \left[ \Phi_{1010} - \Phi_{0101} -
                i \left( \Phi_{1001} - \Phi_{0110} \right) \right]
} \eqno (19)
$$
for (the doublet) $E_{1100}$. Equation (17) illustrates the 
fact that, when going from $q=1$ to $q \ne 1$, the spectrum is 
not only shifted but also split (except for the ground energy 
level). From (17), we obtain the splitting 
$$
\Delta_q \; = \; E_{1100} - E_{2000} \; = \; {3 \over 16} \;
E_0 \; (q-1)^2 \qquad {\hbox {for}} \qquad q - 1 \approx 0. 
\eqno (20)
$$
We note that by taking $q = 1.004$, we get $\Delta_q = 0.33$ 
cm$^{-1}$. The so-obtained value for $\Delta_q$ has the order 
of magnitude of the $(2p \ ^2P_{3 \over 2}) - 
                     (2s \ ^2S_{1 \over 2}, \ 
                      2p \ ^2P_{1 \over2})$ 
experimental fine structure splitting.

\smallskip

\noindent {\bf 4. Closing remarks}

\noindent The application of quantum groups to a large class of 
physical problems requires the $q$-deformation of the usual 
dynamical systems. In this direction, the $q$-deformation of 
the one-dimensional harmonic oscillator (Macfarlane 1989, 
Biedenharn 1989) has been a decisive step. The $q$-analogues of 
the hydrogen atom in $\grt$ introduced in this work represent a 
further important step. Part of this work might be extended to
the hydrogen atom in $\grc$ by means of the Hurwitz 
transformation associated to the Hopf fibration on spheres $S^7 
\to S^4$ of compact fiber $S^3$ (Lambert and Kibler 1988, Hage 
Hassan and Kibler 1991).

\smallskip

The hydrogen atom in $\grt$ is a basic dynamical system in 
classical and quantum mechanics in the case $q=1$. (Ordinary quantum 
mechanics can be thought to correspond to the limiting case 
$q=1$.) This system is maximally super-integrable (Evans 1990, 
Kibler and Winternitz 1990) with rich 
dynamical invariance and non-invariance algebras. It would be 
interesting to investigate invariance and non-invariance 
algebras of the $q$-deformed hydrogen atom and to see the 
branching rules when going from $q=1$ (ordinary quantum 
mechanics) to $q \ne 1$ ($q$-deformed quantum mechanics). 
Similar remarks apply to the isotropic harmonic oscillator in 
$\grt$, another maximally super-integrable system.

\smallskip

There are several ways to define a $q$-analogue of the hydrogen 
atom. We have been concerned in this article with two of them. 
There are other solutions connected to pairs of non-commuting 
$q$-bosons. The two $q$-deformed hydrogen atoms defined in 
\S 3 and \S 4 by their discrete spectra are different although 
having the same (ordinary) quantum limit (corresponding to 
$q=1$). This constitutes in fact a general problem 
we face when dealing with $q$-deformed objects (cf., for 
example, the case of coherent states introduced by 
Katriel and Solomon (1991)). We presently do not have a simple 
correspondance principle for associating a (unique) $q$-analogue 
to a given mathematical or physical object. The use of a 
clearly defined, via the notion of $q$-derivative, 
$q$-deformed Schr\"odinger equation should lead to some 
standardisation of $q$-deformed objects. 

\smallskip

Among the two $q$-deformed hydrogen atoms discussed above, 
the one obtained via 
the KS transformation is certainly the most interesting. It 
leads to a spectrum which is shifted and split with respect to 
the ordinary quantum limit corresponding to $q=1$. Furthermore, 
the $2s-2p$ Dirac splitting can be reproduced by adjusting the 
value of $q$ and this yields a value close to 1. The latter 
result should not be taken too seriously but it however 
constitutes a model of the $2s-2p$ splitting without relativistic 
quantum mechanics. 

\smallskip

We close this section with a remark about the $q$-bosonisation 
(7) of the spherical angular momentum $\{J_-, J_3, J_+\}$ leading 
to $su_q(2)$. We may also define 
       an hyperbolic angular momentum $\{K_-, K_3, K_+\}$ by 
$$
K_- \; = \; a_+a_- \qquad \quad 
K_3 \; = \; {1\over 2} \left( N_1 + N_2 + 1 \right) \qquad \quad 
K_+ \; = \; a^+_+ a^+_-
\eqno (21)
$$
with the property
$$
K_- |jm> \; = \; {\sqrt {[j-m] [j+m]}} \; |j-1, m> 
$$
$$
K_3 |jm> \; = \; ( j + {1 \over 2} ) \; |jm> \eqno (22) 
$$
$$
K_+ |jm> \; = \; {\sqrt {[j-m+1] [j+m+1]}} \; |j+1,m>.
$$
The operators $K_-$, $K_3$ and $K_+$ satisfy the commutation relations
$$
[K_3, K_{\pm}]_- \; = \; {\pm} \; K_{\pm}  \qquad \quad 
[K_+, K_-]_- \; = \; - \; [2K_3]
\eqno (23)
$$
and thus span the quantum algebra $su_q(1,1)$. There are 
four other bilinear forms of the $q$-bosons $a_+$, $a_+^+$, 
                                         $a_-$ and $a_-^+$
which can be formed in addition to the $J$'s and the $K$'s, 
viz,
$$
k^+_+ \; = \; - \; a^+_+ a^+_+ \qquad 
k^+_- \; =      \; a^+_- a^+_- \qquad
k^-_- \; = \; - \; a_+ a_+     \qquad
k^-_+ \; =      \; a_- a_-.
\eqno (24)
$$
The operators $k$'s are clearly shift operators for the quantum 
numbers $j$ and $m$. It can be proved that the set 
$\left \{ k_+^+, k_-^+, k_-^-, k_+^- \right \}$ span, 
together with the sets 
$\left\{ J_-, J_3, J_+ \right\}$ and 
$\left\{ K_-, K_3, K_+ \right\}$, a 
quantum algebra which may be identified to $so(3,2) \sim sp(4,\gr)$ 
when $q$ goes to 1. The quantum algebra $so_q(3,2)$ 
might be useful for studying 
the Wigner-Racah algebra of the group $SU_q(2)$ as was 
done in the case $q=1$ (Kibler and Grenet 1980). 

\smallskip

\noindent {\bf Acknowledgments}

\noindent One of the authors (M K) thanks Y Saint-Aubin for communicating 
his lectures notes (Saint-Aubin 1990) on quantum groups. He is grateful 
to J Katriel, G Rideau and S~L Woronowicz for interesting discusions.

\vfill\eject
\baselineskip = 0.84 true cm

\noindent {\bf References}

 \noindent Biedenharn L~C 1989 {\it J. Phys. A: Math. Gen.} 
{\bf 22} L873

 \noindent Boiteux M 1972 {\it C. R. Acad. Sci., Paris} {\bf 274B} 867

 \noindent Cornish F~H~J 1984 {\it J. Phys. A: Math. Gen.} 
{\bf 17} 2191

 \noindent Drinfel'd V~G 1985 {\it Sov. Math. Dokl.} {\bf 32} 254

 \noindent Evans N~W 1990 {\it Phys. Lett.} {\bf 147A} 483

 \noindent Hage Hassan M and Kibler M 1991 On Hurwitz 
transformations {\it Report} LYCEN 9110

 \noindent Jimbo M 1985 {\it Lett. Math. Phys.} {\bf 10} 63

 \noindent Katriel J and Solomon A~I 1991 {\it J. Phys. A: Math. Gen.} 
{\bf 24} 2093 

 \noindent Kibler M and Grenet G 1980 {\it J. Math. Phys.} 
{\bf 21} 422

 \noindent Kibler M and N\'egadi T 1983a {\it Lett. Nuovo Cimento} {\bf 37} 225

 \noindent --------- 1983b {\it J. Phys. A: Math. Gen.} {\bf 16} 4265

 \noindent --------- 1984 {\it Phys. Rev.} A {\bf 29} 2891

 \noindent Kibler M, Ronveaux A and N\'egadi T 1986 {\it J. Math. Phys.} 
{\bf 27} 1541

 \noindent Kibler M and Winternitz P 1990 {\it Phys. Lett.} {\bf 147A} 
338

 \noindent Kulish P~P and Reshetikhin N~Yu 1989 {\it Lett. 
Math. Phys.} {\bf 18} 143

 \noindent Kustaanheimo P and Stiefel E 1965 
{\it J. Reine Angew. Math.} {\bf 218} 204

 \noindent Lambert D and Kibler M 1988 {\it J. Phys. A: Math. Gen.} 
{\bf 21} 307

 \noindent Macfarlane A~J 1989 {\it J. Phys. A: Math. Gen.} 
{\bf 22} 4581

 \noindent Nomura M 1990 {\it J. Phys. Soc. Jpn.} {\bf 59} 2345 

 \noindent Pauli W 1926  {\it Z. Phys.} {\bf 36} 336 

 \noindent Quesne C 1991 {\it Phys. Lett.} {\bf 153A} 303

 \noindent Saint-Aubin Y 1990 Quantum groups and their 
application to conformal quantum field theories {\it Report} CRM-1663

 \noindent Schwinger J 1952 On angular momentum {\it Report} 
U.S. AEC NYO-3071 (reprinted in 1965 {\it Quantum Theory of 
Angular Momentum} ed L~C Biedenharn and H van Dam (New York: 
Academic) p~229) 

 \noindent Sun C-P and Fu H-C 1989 {\it J. Phys. A: Math. Gen.} 
{\bf 22} L983

 \noindent Woronowicz S~L 1987 {\it Comm. Math. Phys.} {\bf 111} 
613

\bye